\documentclass{INTERSPEECH2023}

\interspeechcameraready

\usepackage[absolute]{textpos}
\title{Confidence-based Ensembles of End-to-End Speech Recognition Models}
\name{Igor Gitman$^1$, Vitaly Lavrukhin$^1$, Aleksandr Laptev$^{1,2}$, Boris Ginsburg$^1$}
\address{
  $^1$NVIDIA, USA \\
  $^2$ITMO University, Russia}
\email{igitman@nvidia.com, vlavrukhin@nvidia.com, alaptev@nvidia.com, bginsburg@nvidia.com}

\begin{document}

\maketitle

\begin{abstract}
The number of end-to-end speech recognition models grows every year. These models are often adapted to new domains or languages resulting in a proliferation of expert systems that achieve great results on target data, while generally showing inferior performance outside of their domain of expertise. We explore combination of such experts via confidence-based ensembles: ensembles of models where only the output of the most-confident model is used. We assume that models' target data is not available except for a small validation set. We demonstrate effectiveness of our approach with two applications. First, we show that a confidence-based ensemble of 5 monolingual models outperforms a system where model selection is performed via a dedicated language identification block. Second, we demonstrate that it is possible to combine base and adapted models to achieve strong results on both original and target data. We validate all our results on multiple datasets and model architectures.
\end{abstract}

\begin{textblock}{100}(20,290)
\noindent To appear in \emph{Proc. INTERSPEECH 2023, August 20-24, 2023, Dublin, Ireland}
\end{textblock}
\section{Introduction}
\begin{figure}[t]
  \centering
  \includegraphics[width=\linewidth]{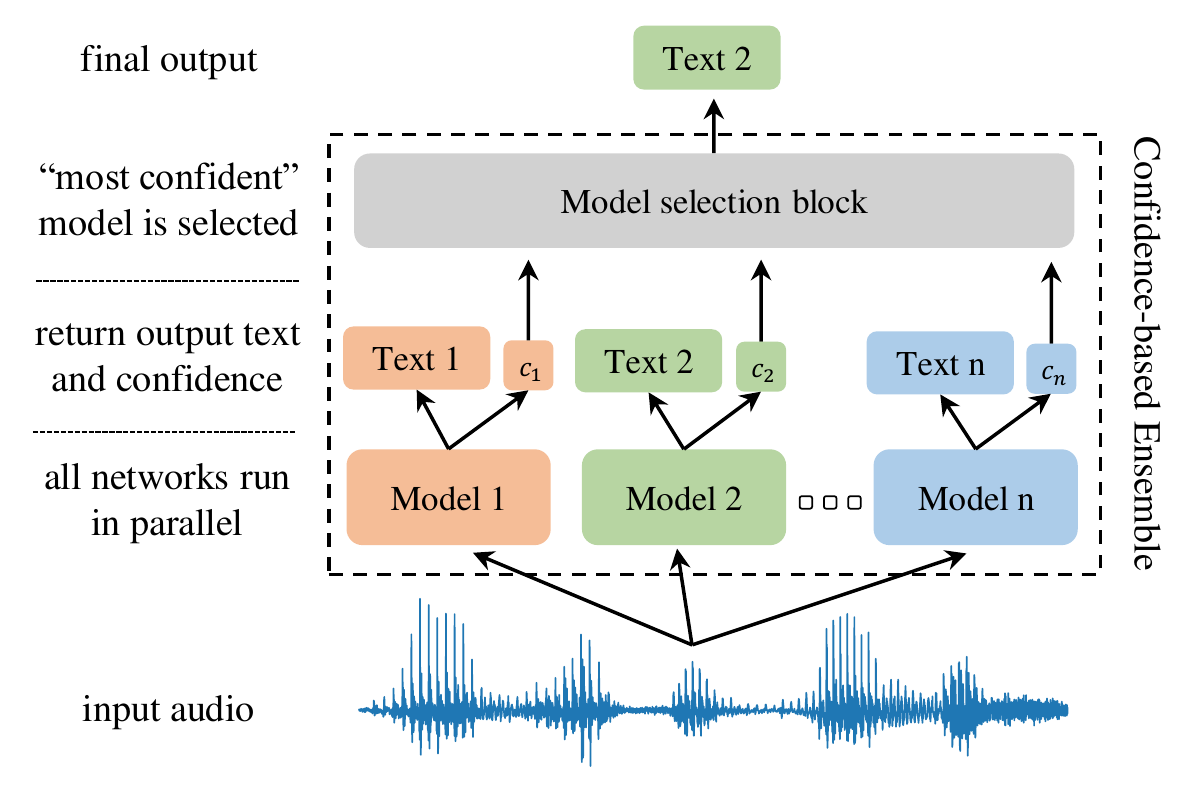}
  \caption{An illustration of a confidence-based ensemble applied to ASR. The same input audio is passed to all models. They run in parallel to produce an output text as well as an estimation of the output's correctness (confidence). A confidence of each model can generally consist of multiple values (e.g., acoustic and language model scores). Model selection block can be implemented as any mapping from confidence values to the index of the ``most confident'' model.}
  \label{fig:conf-ensemble}
\end{figure}

Deep neural network (DNN) models are commonly combined in an ensemble to improve predictions' accuracy. Most widely used types of ensembles include bagging~\cite{opitz1997empirical}, boosting~\cite{schwenk2000boosting} and stacking~\cite{deng2012scalable}. A less popular ensembling approach is to only use an output of a single model that is deemed best for the current input. A typical way to pick the ``best'' output is to select a model with the highest confidence score, which provides an estimate of how likely the output is to be correct. We refer to such systems as \emph{confidence-based ensembles} (see 
Figure~\ref{fig:conf-ensemble}). 

There exists a number of applications of confidence-based ensembles to automatic speech recognition (ASR) and classification. Metze et al.~\cite{metze2000confidence} used confidence scores of multiple monolingual Hidden Markov Model (HMM)-based ASR models to improve language identification (LID) performance. This idea has been later applied to build a hybrid DNN-HMM multilingual ASR model where a combination of a dedicated LID block and a confidence-based model selection was used~\cite{lin2012recognition,gonzalez2014real}. Wang et al.~\cite{wang2019signal} extended this further to include additional acoustic and language model scores and trained a neural network to make the final model selection. Confidence-based ensembles have also been applied to dialectal and accented speech recognition. Soto et al.~\cite{soto2016selection} showed that a confidence-based hypothesis selection improves recognition results on Arabic dialects. Kukk and Alum{\"a}e~\cite{kukk2022improving} improved accented speech LID by directly building text classification models on top of the ASR output.

While there are many applications of confidence-based ensembles of HMM-based ASR systems, there is no study on using confidence to combine multiple \emph{end-to-end} neural ASR models. We fill this gap by building ensembles of connectionist temporal classification (CTC)~\cite{graves_connectionist_2006} and recurrent neural network transducer (RNNT)~\cite{graves2012sequence} models (using Conformer~\cite{gulati2020conformer} as an encoder). There are different ways how to define confidence measure for such models. A traditional approach is to use a maximum probability of the output tokens~\cite{hendrycks2016baseline}. Laptev et al.~\cite{laptev2022fast} observed that using entropy-based measures helps to reduce networks' overconfidence. It is possible to train separate models for confidence estimation~\cite{woodward2020confidence,li2021confidence}, but such methods are not directly applicable to pretrained models, so we do not consider them in this paper. Instead, we focus on the collection of entropy-based measures proposed in~\cite{laptev2022fast} and systematically study their effect on the quality of confidence-based ensembles.

\begin{table*}[t]
    \caption{Comparison of different confidence estimation methods for ensembles of Conformer models. All the numbers in the table represent an average per-dataset accuracy on the validation subsets. ``LID 5 languages'' is a combination of MCV, MLS and VoxPopuli data. For the ``LID 5 languages'' and ``SLR83'' columns we report mean ± standard deviation of results across all datasets. ``untuned max-prob'' confidence is defined as a product of probabilities of emitted tokens (including blanks). ``default confidence'' corresponds to Rényi entropy with linear normalization, mean aggregation (excluding blanks), $T = 1.0, \alpha = 0.25$. ``tuned confidence'' is the best confidence measure, which is generally different for each dataset.} 

    \centering
    \begin{tabular}{lccc|ccc}
    \toprule
    {} &   \multicolumn{3}{c|}{\textbf{Conformer-Transducer}} & \multicolumn{3}{c}{\textbf{Conformer-CTC}} \\
    \textbf{Confidence}& \textbf{LID 5 languages} & \textbf{CORAAL} & \textbf{SLR83} & \textbf{LID 5 languages} & \textbf{CORAAL} & \textbf{SLR83}               \\
    \midrule
    untuned max-prob   &  98.18 ± 0.77 &  88.00 &  75.32 ± 4.35 &  98.76 ± 0.82 &  84.89 &  57.84 ± 4.53 \\
    default confidence &  99.13 ± 0.44 &  94.13 &  81.85 ± 4.73 &  99.37 ± 0.39 &  91.39 &  77.08 ± 5.23 \\
    tuned confidence   &  99.39 ± 0.32 &  95.31 &  93.60 ± 1.88 &  99.48 ± 0.37 &  93.31 &  90.37 ± 4.35 \\
    \bottomrule
    \end{tabular}  
    \label{tab:conf-comp}
\end{table*}

We use two applications to evaluate the effectiveness of our approach. First, we consider a task of multilingual ASR. The current trend in building multilingual end-to-end ASR models is to train a single network that can recognize multiple languages~\cite{punjabi2021joint,li2022language,zhou2022configurable}. Another popular line of work is to build mixture-of-expert systems that use separate sub-networks specializing in different languages~\cite{gaur2021mixture,mole}. However, such techniques require significant computational resources for training and might produce worse recognition than monolingual predictors~\cite{li2022language}. An alternative approach is to combine monolingual models into a multilingual system using a dedicated language identification (LID) block~\cite{gonzalez2014real}. We show that confidence-based ensembles outperform dedicated LID models on long audio segments ($>5$ seconds) and can be combined with LID model's scores to improve predictions on short segments. 

We then show that it is possible to use confidence-based ensembles for the models sharing the same input language. We consider a task of accent and dialect adaptation. A typical solution to this problem is to finetune an ASR model on the target data. A common issue of this approach is that model's performance on the original domain might significantly degrade, which is known as catastrophic forgetting~\cite{french1999catastrophic}. While many techniques have been proposed to address this issue~\cite{zhao2021unified,hwang2022large}, they often require access to the original training data, which might not be available. Majumdar et al.~\cite{majumdar2023damage} proposes using limited training strategy and regularized adapter modules to reduce the degradation on the source domain without access to the original training data. In this paper we show that a simple confidence-based combination of original and finetuned models achieves significantly better results. Additionally, we demonstrate that the proposed ensemble can interpolate between better performance on source or target domain via a \emph{runtime} change in the weights of the confidence scores.



A clear limitation of most ensemble systems is that required computation grows linearly with each added model. To address this problem, we show that it is possible to use an output of intermediate layers for confidence estimation. We demonstrate this technique through an ensemble of models finetuned with intermediate CTC loss~\cite{lee2021intermediate} and show that using only 4 out of 18 layers is enough for accurate model selection. We apply this technique to both CTC and Transducer models and find that it generally improves the base models' accuracy.

Summing up, our findings are as follows:
\begin{itemize}
    \item Confidence-based ensembles of end-to-end ASR models have better recognition accuracy than a system with a dedicated LID block on long audio segments. Our method provides approximately $10\%$ relative word-error-rate reduction (WERR) compared to the state-of-the-art LID systems on 3 public multilingual datasets.
    \item Our method can be used to significantly limit catastrophic forgetting during model adaptation. Compared to the constraint adaptation of~\cite{majumdar2023damage}, confidence-based ensembles show $10$--$50\%$ WERR on the target data, while being $2\%$ relatively better on the source domain.
    \item Confidence can be reliably estimated from the output of early layers (4 out of 18) if the model was trained with intermediate CTC loss~\cite{lee2021intermediate}. This reduces the runtime cost of adding a new model to the ensemble by $4.5$ times.
\end{itemize} 

\section{Method}
An illustration of confidence-based ensembles is presented in Figure~\ref{fig:conf-ensemble}. This general structure allows for any vector of confidence values to be produced by each model and any non-linear mapping of the confidence vectors to the model indices to be used as a ``model selection block''. In our experiments we adopt a simpler pipeline. We only use a single confidence score for each model and train a logistic regression (LR) on a few audio samples to predict matching model index from the generated confidence values.

\subsection{Model selection}
Formally, our model selection block is defined as follows. Let's denote the set of models in the ensemble as $\mathcal{M} = \{m_k\}_{k=1}^M$ and the set of evaluation datasets as $\mathcal{D} =\{d_i\}_{i=1}^D$. We assume that there is a surjective mapping $L: \mathcal{D} \rightarrow \mathcal{M}$ that defines which model is considered ``correct'' for which dataset. We select a small set of $N$ utterances\footnote{We use $N=100$ in our experiments, but for datasets with a lot of variability it might be better to use larger training data.} of the \emph{training} subset of each dataset $d_i$ denoted as $\{t_i^j\}_{j=1}^{N}$. Finally, let's define $c_k(t_i^j)$ as the confidence score of the model $k$ computed on the sample $j$ from the dataset $i$. Then we train $\text{LR}: \mathcal{R}^M \rightarrow \mathcal{M}$ to map $x_i^j = [c_1(t_i^j), c_2(t_i^j), ..., c_M(t_i^j)] \rightarrow y_i^j = L(d_i)$.

Recall that logistic regression is a linear classifier that only needs to train $n + 1$ coefficients for $n$ models. E.g., in the case of a 2-model ensemble, LR produces the following model selection rule: $ac_1 + bc_2 > c$. While a more powerful non-linear functions might achieve better results, using a simple LR allows us to train model selection block with only a few audio samples and helps to avoid overfitting.

\subsection{Evaluation metrics}
To evaluate the accuracy of the ASR models we use a traditional word-error-rate (WER) metric. To evaluate the accuracy of the model selection within ensemble we use the \emph{average per-dataset accuracy ($A_{\text{avg}}$)} metric:
\begin{equation*}
    A_{\text{avg}} = \frac{1}{D}\sum_{i=1}^D \frac{1}{D_i}\sum_{j=1}^{D_i} \left[\text{LR}\left(c_1(v_i^j), ..., c_m(v_i^j)\right) = L\left(d_i\right)\right]
\end{equation*}
\noindent
where $v_i^j, j=1..D_i$ are the \emph{validation/test} utterances from dataset $d_i$ and $D_i$ is the size of the validation/test set. $[X = Y]$ is an indicator function that equals $1$ if $X = Y$ and $0$ otherwise. Put simply, an average per-dataset accuracy is a mean of the model prediction accuracies across all datasets. We use this metric instead of a regular multi-class prediction accuracy in order to account for label imbalances present in some dataset combinations. E.g., multilingual datasets can be very imbalanced, but we assume that the performance on each language is of equal importance.

\section{Experiments}
 For all experiments we followed the same evaluation setup. We fit LR on 100 training utterances and tuned LR's hyperparameters (regularization strength and class weights in case of imbalanced datasets) maximizing the $A_{\text{avg}}$ metric on the validation subsets of all datasets. In the following tables we report the results on the test subsets unless noted otherwise. All experiments were performed using NeMo toolkit\footnote{\url{https://github.com/NVIDIA/NeMo}}~\cite{kuchaiev2019nemo}.

\subsection{Datasets}
\subsubsection{Multilingual ASR.}
To evaluate the performance of our method we used 3 public multilingual ASR datasets: Mozilla Common Voice (MCV)~\cite{ardila2020common}, Multilingual LibriSpeech (MLS)~\cite{Pratap2020} and VoxPopuli~\cite{wang2021voxpopuli}. For all datasets we used standard train, validation and test splits. We refer the reader to the original papers for additional information about the used datasets. We built ensembles that supports five languages: English, Italian, Spanish, German and French. For all languages we used ``Conformer Large'' models from NVIDIA NGC catalog.

\subsubsection{Accent and dialect adaptation.}

We tested our method on 2 public datasets: Corpus of Regional African American Language (CORAAL)~\cite{kendall2018corpus} and Open-source Multi-speaker Corpora of the English Accents in the British Isles (SLR83)~\cite{demirsahin2020open}. We manually split CORAAL into train, validation and test splits containing 18, 3 and 22 hours respectively. The training, validation and test sets contained audio from different speakers. For SLR83 we re-used data splits from~\cite{majumdar2023damage}. To be able to have direct comparison with~\cite{majumdar2023damage} we used Conformer Large models trained on LibriSpeech dataset~\cite{panayotov2015librispeech} as the base for model adaptation on SLR83 datasets. For CORAAL, we started finetuning from Conformer Large trained on 24325 hours of English speech from NVIDIA NGC catalog.

\subsection{Impact of the confidence measure}
In this section we study the effect of choosing a confidence estimation method on the ensemble's quality. We evaluated Gibbs, Tsallis~\cite{tsallis1988possible} and Rényi~\cite{renyi1961measures} entropies (with both linear and exponential normalization) proposed in~\cite{laptev2022fast} as well as the maximum probability as confidence measures. To aggregate confidence information across time steps we tested minimum, maximum, mean and product aggregation functions. Both RNNT and CTC models use a special \textit{blank} symbol to align input to the output. We test both including and ignoring  blanks  in the confidence aggregation. For all measures we tuned the softmax's temperature  $T \in [0.01, 0.05, 0.1, 0.25, 0.5, 0.75, 1.0, 2.0, 5.0, 10.0]$. For the entropy-based measures we also ran a grid-search over the parameter $\alpha \in [0.1, 0.2, 0.25, 0.33, 0.5, 1.0]$. Overall, the search space spans 2960 combinations of confidence measures.

While the best performing measure for each dataset is typically different, we found that using \emph{Rényi entropy with linear normalization, mean aggregation, $T = 1.0, \alpha = 0.25$ and blank symbols excluded} tends to perform well across all evaluated tasks\footnote{This measure might not be well-calibrated as it is tuned to optimize the model-selection accuracy.}. Table~\ref{tab:conf-comp} shows the comparison of using the product of probabilities of the output tokens (untuned max-prob), Rényi entropy with the aforementioned parameters (default confidence) and the best confidence measure for each dataset (tuned confidence). The proposed default confidence always outperforms maximum probability and can be further improved if confidence hyperparameters are tuned for each dataset.

Due to the space constraints all subsequent results are reported with default confidence and only for Transducer models.

\begin{table}[t!]
  \caption{LID accuracy and WER on VoxPopuli, MLS and MCV datasets. All numbers show an average of the results across all 5 languages. Confidence-based ensembles outperform all baseline LID systems.}
  \centering
    \begin{tabular}{lrrr}
    \toprule
    \textbf{LID model} &      \textbf{VoxPopuli} &     \textbf{MLS} &     \textbf{MCV} \\
    \midrule
    \multicolumn{4}{l}{\emph{LID accuracy}}   \\
    ECAPA-TDNN-CE~\cite{desplanques20_interspeech}   & 96.29 &  98.41 & 94.42       \\    
    XLS-R~\cite{babyxlsr2022_interspeech}            & 97.63 &  98.69 & 97.47       \\
    confidence-based                                 & \textbf{98.76} &  \textbf{99.69} & \textbf{98.82}      \\
    \midrule
    \multicolumn{4}{l}{\emph{WER}} \\
    oracle                                           &  9.44  &  5.89 &  6.05        \\
    ECAPA-TDNN-CE~\cite{desplanques20_interspeech}   & 11.26  &  6.90 &  9.44         \\    
    XLS-R~\cite{babyxlsr2022_interspeech}            & 10.53  &  6.78 &  7.23         \\
    confidence-based                                 &  \textbf{9.57}  &  \textbf{6.04} &  \textbf{6.24}          \\
    \bottomrule
    \end{tabular}
  \label{tab:voxpopuli-results}
\end{table}

\begin{table}[t!]
  \caption{Model selection accuracy with different audio duration on VoxPopuli dataset. Only utterances with more than 15 seconds of audio were used to have the same evaluation set for all durations. ``Combination'' column shows that using a combination of confidence and LID scores can improve the results for short segments.}
  \centering  
    \begin{tabular}{lrrrr}
    \toprule
    \textbf{Audio duration} &  \textbf{Confidence} & \textbf{XLS-R} & \textbf{Combination}  \\
    \midrule
    3 sec      &  90.63 &  93.45 &  96.00 \\
    5 sec      &  95.50 &  95.45 &  97.56 \\
    10 sec     &  98.05 &  97.07 &  98.86 \\
    15 sec     &  99.02 &  97.56 &  99.46 \\
    full audio     &  99.57 &  97.67 &  99.57 \\
    \bottomrule
    \end{tabular}
  \label{tab:duration}
\end{table}

\begin{table*}[t!]
  \caption{Comparison of the WER for different model adaptation techniques on the SLR83 dataset. The LibriSpeech test-other WER is averaged across all speakers. We abbreviate each speaker id using the first two letters of speaker's dialect and the first letter of gender (e.g., ``Ir-M'' stands for ``Irish-English Male''). Our ``base'' model is the same as in~\cite{majumdar2023damage}, but the ``finetuned'' model is generally better on the target data as we do not constraint training.}
  \centering  
    \begin{tabular}{lcrrrrrrrrrrr}
    \toprule
    \textbf{Model} &           \textbf{LS other} &   \textbf{Ir-M} &  \textbf{Mi-F} &   \textbf{Mi-M} &   \textbf{No-F} &   \textbf{No-M} &   \textbf{Sc-F} &   \textbf{Sc-M} &  \textbf{So-F} &   \textbf{So-M} &  \textbf{We-F} &   \textbf{We-M} \\
    \midrule
    base                  &         5.12 &  20.69 &  9.61 &  11.25 &  11.11 &  10.18 &  12.26 &  11.94 &  9.70 &  10.22 &  8.51 &  11.46 \\
    finetuned             &  7.85 ± 2.28 &   9.17 &  7.51 &   8.36 &   7.60 &   6.87 &   8.32 &   6.35 &  4.14 &   3.59 &  4.67 &   6.49 \\
    \midrule 
    constrained~\cite{majumdar2023damage} &  5.40 ± 0.05 &  15.86 &  8.40 &   9.43 &   9.33 &   8.54 &  10.00 &   8.68 &  7.73 &   7.90 &  6.64 &   9.70 \\
    ensemble    &  \textbf{5.26 ± 0.11} &   \textbf{9.17} &  \textbf{7.51} &   \textbf{8.36} &   \textbf{7.63} &   \textbf{6.78} &   \textbf{8.40} &   \textbf{6.82} &  \textbf{4.64} &   \textbf{3.98} &  \textbf{4.79} &   \textbf{6.70} \\
    \bottomrule
    \end{tabular}
    
  \label{tab:slr83}
\end{table*}

\begin{table}[t]
  \caption{WER after model adaptation on CORAAL dataset. We trained ensemble to select ``base'' model for both LibriSpeech and VoxPopuli datasets and ``finetuned'' model for CORAAL dataset. The last two rows demonstrate capability of ensemble to prioritize original vs target domain via a runtime adjustment of logistic regression probability threshold.}
  \centering  
    \begin{tabular}{lrrr}
    \toprule
    \textbf{Model} &  \textbf{LS other} &  \textbf{VoxPopuli} &  \textbf{CORAAL} \\
    \midrule
    base                      &      3.73 &       6.37 &   31.23 \\
    finetuned                 &      5.25 &      10.51 &    8.29 \\
    \midrule
    ensemble default       &      3.82   &    6.57  &  8.72 \\
    tuned for base    &      3.76   &    6.39  &  9.77 \\
    tuned for target  &      4.31   &    8.85  &  8.33 \\
    \bottomrule
    \end{tabular}
  \label{tab:coraal}
\end{table}

\subsection{Multilingual ASR}
\label{exp:multilingual-asr}
In this section we evaluate the performance of the confidence-based ensembles for the multilingual ASR task. We compared our method with several state-of-the-art LID models~\cite{desplanques20_interspeech,babyxlsr2022_interspeech}. Since these models were trained to recognize much larger set of languages, we constrained them to only select the highest probability index out of the 5 evaluated languages. Table~\ref{tab:voxpopuli-results} shows LID accuracy and WER of the resulting systems on VoxPopuli, MLS and MCV datasets. Table~\ref{tab:duration} shows how the LID accuracy changes for different audio duration. Our experiments demonstrate that while confidence-based ensembles outperform dedicated LID models on the long audio segments ($>5$ seconds), they underperform the best baseline for shorter durations. Last column of Table~\ref{tab:duration} shows that using a combination of confidence and LID scores as an input to logistic regression obtains significantly better predictions for the short segments.


\subsection{Accent and dialect adaptation}
For all finetuning experiments in this section we ran training for 100 epochs on a single 16GB NVIDIA V100 GPU re-using training configuration of the base model, except for the batch size and learning rate parameters. Batch size was selected to fully utilize GPU memory and learning rate was tuned over 8 log-uniform grid points in $[10^{-7}, 10^{-3}]$. Table~\ref{tab:slr83} shows the comparison of the confidence-based ensembles with the constraint adaptation approach of~\cite{majumdar2023damage} on the SLR83 dataset. Our method achieves significantly better accuracy on both original and target domains. Note that for one dialect confidence-based ensemble outperforms the finetuned model. This is possible because the base model can be ``incorrectly'' selected on utterances where it has lower WER than the finetuned model.

Table~\ref{tab:coraal} shows the results of the confidence-based ensemble after adaptation on the CORAAL dataset. The last two rows of the table demonstrate an ability of the ensemble to trade-off between performance on the target and the original domain via a runtime change of the logistic regression threshold. Based on the specific requirements users can adjust the performance of the model with a simple configuration change.

\subsection{Decreasing runtime cost}
A clear limitation of model ensembles is that runtime cost increases linearly with ensemble size. To partially overcome this, we propose to use outputs of the intermediate layers for model selection\footnote{This approach requires finetuning models and thus can only be used if the training data is available.}. Table~\ref{tab:interctc} shows that it is possible to get a high-quality confidence estimate using intermediate encoder layers. We finetuned all models for 100 epochs on the combination of MCV, MLS and VoxPopuli datasets using intermediate CTC loss~\cite{lee2021intermediate} applied to different layers and used the output of those layers to compute confidence scores. Following the original paper, we added the new loss with coefficient of $0.3$. We re-used training configuration of the base models except the learning rate which was 10 times smaller. Unlike the original paper we applied this technique to both CTC and Transducer models and it generally increased models' accuracy because of the extra regularization.
Surprisingly, even though WER of intermediate layers is significantly worse, using intermediate confidence has only a minor reduction in the model-selection accuracy. 

\begin{table}[t]
  \caption{Model selection accuracy using confidence estimation from different layers. All models have been finetuned with intermediate CTC loss~\cite{lee2021intermediate}.}
  \centering  
    \begin{tabular}{lrrr}
    \toprule
    \textbf{Confidence from} &      \textbf{CORAAL} &     \textbf{VoxPopuli} &     \textbf{MLS} \\
    \midrule
    Layer 4               &  91.58 &  98.67 &  99.80 \\
    Layer 9               &  93.32 &  98.08 &  99.75 \\
    Layer 18 (final)      &  94.13 &  98.76 &  99.69 \\
    \bottomrule
    \end{tabular}
  \label{tab:interctc}
\end{table}

\section{Conclusion}
In this paper we showed that confidence-based ensembles of end-to-end ASR models can improve state-of-the-art results for several speech recognition problems. Our method achieves close to $10\%$ WERR on 3 multilingual ASR datasets compared to systems using a dedicated LID block. It also improves results on accent and dialect adaptation tasks by $10$--$50\%$ WERR.

However, there are several limitations that need to be acknowledged. Confidence-based ensembles are not well suited for latency-critical applications as they require a few seconds of audio to select the most confident model. The runtime cost grows linearly with each added model, which limits the practically useful ensemble size. Finally, given enough compute and data, it is likely possible to build specialized models that would outperform confidence-based ensembles on most tasks.

Taking these limitations into account, we think that confidence-based ensembles can be useful in a wide range of applications beyond what is covered in our experiments. We consider building confidence-based ensembles as a general technique to combine multiple black-box expert models into a single system that achieves competitive results on all target domains. The models can be combined with almost no additional training and without the need to share either data or model weights. We think that these properties can enable new applications of speech technology for users who don't have access to large compute clusters or big training datasets.

\bibliographystyle{IEEEtran}
\bibliography{mybib}

\end{document}